# Thermal Annealing and Radiation Effects on Structural and Electrical Properties of NbN/GaN Superconductor/Semiconductor Junction

Stephen Margiotta[1, 2], Binzhi Liu[3], Saleh Ahmed Khan[1], Gabriel Calderon Ortiz[3], Ahmed Ibreljic[1], Jinwoo Hwang[3], A F M Anhar Uddin Bhuiyan[1, a]

[1]Department of Electrical and Computer Engineering, University of Massachusetts Lowell, Lowell, Massachusetts 01854, USA

[2]Lincoln Laboratory, Massachusetts Institute of Technology, Lexington, Massachusetts 02421, USA

[3]Department of Materials Science and Engineering, The Ohio State University, Columbus, Ohio 43210, USA

[a]Corresponding author Email: anhar_bhuiyan@uml.edu

## Abstract

In the rapidly evolving field of quantum computing, niobium nitride (NbN) superconductors have emerged as integral components due to their unique structural properties, including a high superconducting transition temperature ($T_c$), exceptional electrical conductivity, and compatibility with advanced device architectures. This study investigates the impact of high temperature annealing and high dose gamma irradiation on the structural, electrical and superconducting properties of NbN films grown on GaN via reactive DC magnetron sputtering. The as-deposited cubic δ-NbN (111) films exhibited high intensity distinct x-ray diffraction (XRD) peak, a high $T_c$ of 12.82 K, and an atomically flat surface. Annealing at 500 °C and 950 °C for varying durations revealed notable structural and surface changes. High resolution scanning transmission electron microscopy (STEM) indicated improved local ordering, while atomic force microscopy (AFM) showed reduced surface roughness after annealing. X-ray photoelectron spectroscopy (XPS) revealed a gradual increase in the Nb/N ratio with higher annealing temperatures and durations. High-resolution XRD and STEM analysis showed lattice constant modifications in δ-NbN films, attributed to residual stress changes following annealing. Additionally, XRD φ-scans revealed sixfold symmetry in NbN films due to rotational domains relative to GaN. While $T_c$ remained stable after annealing at 500 °C, increasing the annealing temperature to 950 °C degraded $T_c$ to 8.7 K and reduced the residual resistivity ratio (RRR) from 0.85 in as-deposited films to 0.29 after 30 minutes annealing. The effects of high-dose gamma radiation (5 Mrad (Si)) were also studied, demonstrating minimal changes to crystallinity and superconducting performance, indicating excellent radiation resilience. These findings highlight the potential of NbN superconductors for integration into advanced quantum devices and its suitability for applications in radiation-intensive environments such as space, satellites, and nuclear power plants.

## I. INTRODUCTION

Niobium nitride (NbN) has long been a material of significant interest in the field of superconductivity, primarily due to its remarkable superconducting transition temperature ($T_c$), which typically ranges from 15 to 17 K [1-14]. The superconducting properties of NbN have been a major focus of research for several decades, significantly contributing to the advancement of cutting-edge technologies such as superconducting quantum computers [15], single-photon detectors [16,17], and hot electron bolometers [18,19]. The high $T_c$ of NbN, coupled with its robust structural and electrical characteristics, also makes it a leading candidate for applications that demand superior performance under extreme conditions. In addition to a high Tc, NbN exhibits other key superconducting properties that make it highly suitable for quantum technologies. Notably, NbN possesses a high upper critical magnetic field [20], high kinetic inductance [21, 22], and a large ratio of London penetration depth to coherence length [23]. These properties are crucial in applications such as superconducting resonators and single flux quantum (SFQ) electronics. The cubic δ-phase of NbN, in particular, is known for yielding the highest $T_c$ values [24], highlighting the critical need to control its structural characteristics such as crystal structure, lattice constant, and nitrogen content- to maximize its superconducting potential. Precise regulation of these parameters is crucial for effectively integrating NbN into quantum devices and leveraging its exceptional properties.

NbN thin films have been commonly grown using various deposition techniques, including sputter deposition [2,25,26], pulsed laser deposition [27], atomic layer deposition [28], chemical vapor deposition [4], and molecular beam epitaxy (MBE) [9] on different substrates including GaN, SiC, AlN, Sapphire, and $SrTiO_3$. The integration of NbN with group-III nitride semiconductors, such as gallium nitride (GaN) and aluminum nitride (AlN), presents exciting opportunities for the development of hybrid devices that combine the advantages of both superconductors and semiconductors. The relatively small lattice mismatches between the cubic δ-NbN and III-nitrides [26], approximately -0.2% with AlN and -2.7% with GaN, make this combination particularly appealing for epitaxial growth [29,30]. Such structural compatibility facilitates the potential for integrating NbN superconductors with nitride semiconductors on a single wafer, enabling the development of novel devices with enhanced functionality. Previous successes with AlGaN/GaN high electron mobility transistors (HEMTs) on NbN films, which exhibited negative differential resistance below $T_c$ [30], suggest a promising future for combining nitride semiconductors and superconductors in Josephson junctions. Recent advances in molecular beam epitaxy (MBE) have demonstrated NbN films on GaN with $T_c$ values exceeding 10 K [31]. Additionally, the development of NbN-gated AlGaN/GaN HEMTs highlights the potential of NbN/GaN transistor technology for low-noise cryogenic amplifiers in future quantum computing systems [32]. The integration of NbN films on AlN-buffered 300 mm Si wafers also underscores the potential for fabricating large-scale quantum devices [33]. Despite these promising advancements, the growth mechanisms of NbN on nitride semiconductors and their comprehensive characterization remain limited. One critical aspect that has not been fully investigated is the effect of high temperature annealing and high dose radiations on the structural and electrical properties of NbN films deposited on III-nitrides. Annealing plays a crucial role in enhancing film crystallinity, aligning lattice structures, and optimizing the superconducting properties of NbN films [34]. Given the inherent lattice mismatches between NbN and GaN, proper annealing conditions are essential for strain relaxation in the NbN films, leading to improved superconducting performance. Furthermore, understanding the resilience of NbN to radiation is especially important for applications in radiation-hardened environments. While the response of NbN to proton and neutron irradiation has been explored in recent studies [35,36], the effects of gamma ($\gamma$) radiation on its structural and superconducting properties remain unexplored. This is particularly important for space, nuclear and other radiation-intensive environments, where materials are exposed to high doses of gamma radiation, making it crucial to understand how NbN performs under such conditions.

In this study, we aimed to address the gap in understanding by systematically investigating how annealing and radiations affects the structural, electrical and superconducting properties of crystalline NbN superconductors deposited on GaN via reactive DC magnetron sputtering. We explored how variations in annealing temperature and duration influenced the crystallinity, lattice constant, and superconducting transition temperature of the films by comprehensive characterizations. Additionally, we examined the effects of gamma radiation on these properties to understand the properties of NbN in harsh radiation environments. Scanning transmission electron microscopy (STEM) was conducted to evaluate the atomic structure of the films and to examine how high-temperature annealing influences their structural quality. By optimizing annealing conditions, we sought to enhance the structural and superconducting properties of NbN films, making them suitable for integration with nitride semiconductors in advanced quantum devices and ensuring their durability under gamma radiation exposure.

## II. EXPERIMENTAL DETAILS

100 nm thick NbN thin films were deposited on (0001) GaN/sapphire template at room temperature using an Endura 5500 DC Magnetron Sputtering system in an Ar/$N_2$ environment. The deposition was carried out using a 13-inch diameter Nb target (99.95% purity), biased with a DC power of 1500 W, resulting in a deposition rate of 50 nm per minute. The stoichiometry of the NbN was controlled by tuning the Ar/$N_2$ flow ratio to 2.25. Exploration of deposition conditions and process parameters are available in Ref. [25]. The as-deposited NbN films underwent rapid thermal annealing (RTA) at two different temperatures, 500 °C and 950 °C, for 15 seconds, 1 minute, or 30 minutes in an Ar environment. The temperature ramp rate was 50 °C/sec for both annealing temperatures. Surface roughness was measured both before and after annealing using a Bruker atomic force microscope (AFM). The crystalline structure, film quality, and strain were evaluated using XRD measurements, performed with a Rigaku SmartLab X-ray diffraction tool equipped with a Cu Kα radiation source (λ = 1.5418 Å). The temperature dependence of the resistivity of the NbN films was investigated through a four-point probe method. X-ray photoelectron spectroscopy (XPS) measurements were carried out using a Thermo Scientific XPS spectrometer with a monochromatized Al Kα x-ray source ($E_{photon}$ = 1486.6 eV) to determine the Nb and N content in the films. Aberration-corrected Thermo Fisher Scientific Themis-Z scanning transmission electron microscopy was used to obtain high-angle annular dark field (HAADF)

STEM images under 300 kV. TEM samples were milled by the traditional focus ion beam milling with FEI Helios NanoLab 600 DualBeam system. Film thicknesses were confirmed from cross-sectional scanning electron microscopy (SEM) and STEM-HAADF images.

**III. RESULTS AND DISCUSSION**

Figure 1(a) presents the XRD ω-2θ scans for the as-deposited and annealed (500 °C and 950 °C) NbN samples with different annealing durations, indicating strong and distinct peaks of crystalline cubic δ-NbN (111) along with GaN (002). Previous studies have shown that cubic NbN grown on c-axis oriented GaN [31], and other hexagonal substrates, such as 6H-SiC [33,34], and AlN [26], aligns its [111] lattice vector with the out-of-plane c-axis of the substrate due to structural similarities between rock-salt (δ-phase) NbN and wurtzite III-nitrides. Notably, the peak position of the NbN diffraction slightly moves towards higher 2θ angle as the annealing temperature and duration varied as shown in Figure 1(a). The as deposited (111) NbN film shows a peak position around 35.07°, which shifts to between 35.47° and 35.55° after annealing, depending on the annealing temperature and duration. The peak shift is more pronounced under 950 °C annealing conditions compared to those at 500 °C. Although one possible explanation for the peak shift of δ-NbN (111) to a higher angle could be the transformation of the NbN crystal structure into other phases, no such phase modification was observed from atomic resolution STEM imaging as discussed in later paragraphs, indicating that the shift of the peak positions is likely due to residual stress in the NbN films constrained by the underlying GaN substrates. To assess the crystallinity of NbN films, X-ray rocking curves (XRCs) for symmetric (111) diffractions were measured. Figure 1(b) shows the full width at half maximum (FWHM) of the XRCs for as-deposited and annealed samples, indicating strong influence of the annealing temperature and duration on the FWHMs. NbN films annealed at 500 °C exhibited slightly narrower FWHM as compared to the as-deposited film, indicating improved crystallinity. However, films annealed at 950 °C showed larger FWHM values, suggesting that higher annealing temperatures lead to increased mosaicity of the NbN. The lattice orientation of the NbN film relative to the GaN substrate is also analyzed by performing XRD φ-scanning around asymmetrical (105) GaN and (311) NbN planes as shown in Figure 1(c). φ alignment indicates a six-fold symmetry of the NbN due to the presence of two rotational domains, which is consistent with previous report on MBE grown NbN film on GaN [31]. Asymmetric X-ray reciprocal space mapping (RSM) was also conducted to confirm the crystalline structure and strain state of NbN films grown on GaN. Figures 2(a)-(c) show the reciprocal space maps around GaN (11$\bar{2}$4) for as-deposited and annealed NbN films, indicating distinct and strong intensity reciprocal lattice points from (240) δ-NbN plane. The δ-NbN films are found to grow with the growth-axis orientation δ-NbN [111]//GaN [001] and the in-plane orientations δ-NbN [$\bar{1}$10]//GaN [100], as confirmed by STEM analysis in later paragraphs. However, due to the six-fold symmetry of the GaN substrate and the threefold symmetry of cubic δ-NbN along the growth axis, the δ-NbN lattice can align in two possible in-plane orientations as observed from the XRD φ-scan in Figure 1(c), differing by a 60° rotation around the growth axis [31, 37]. Additionally, the NbN films grown on GaN exhibited lattice relaxation, as indicated by the non-coincident in-plane lattice constants ($Q_x$) of NbN peak with those of GaN peak for all three samples.

The surface morphology analysis, as shown in Figures 3 (a)-(g), reveals strong variations in RMS roughness for the NbN films as a function of annealing temperature and duration. The as-deposited NbN film exhibits an RMS roughness of 1.29 nm. Upon annealing, the RMS roughness trends indicate an interplay between annealing temperature, duration, and film stoichiometry. At 500°C, annealing for extended durations (up to 30 minutes) leads to a clear reduction in RMS roughness from 1.29 nm (as-deposited) to 0.60 nm (annealed). This smoothening effect can be attributed to the thermal activation of surface diffusion processes, where atoms on the surface are able to rearrange into more energetically favorable configurations. The reduction in surface roughness is a clear indicator of enhanced surface mobility and atomic rearrangement, leading to improved surface morphology and crystallinity, as also corroborated by the STEM results. Interestingly, annealing at 950°C initially follows a similar trend, reducing surface roughness. However, when annealing duration is extended to 30 minutes, the RMS roughness increases significantly to 2.01 nm. This roughening at higher temperatures over extended durations is likely linked to the degradation of film stoichiometry, particularly the significant loss of nitrogen from the NbN lattice. At elevated temperatures, nitrogen out-diffusion becomes more pronounced, as confirmed by the XPS measurements in Figure 4, which show a substantial increase in the Nb/N ratio from 1.18 at 500°C to 1.83 at 950°C, are likely responsible for the observed increase in surface roughness, as the non-stoichiometric films become less dense and more prone to surface undulations and roughness. The XPS results as discussed in later paragraph further highlight this stoichiometric imbalance, with nitrogen deficiency likely contributing to the increase in electrical resistivity observed at higher annealing temperatures as discussed later.

In the XPS analysis of NbN thin films, as shown in Figure 4, the elemental composition and stoichiometry were determined by examining the core levels of Nb and N after Ar sputter cleaning of the air-exposed surface. XPS spectra of Nb 3s and N 1s core levels were measured for both as-deposited and annealed samples (500°C and 950°C). For the as-deposited film, the Nb/N ratio was 1.13, indicating a slight deviation from stoichiometry. Upon annealing at 500°C, the Nb/N ratio increased modestly to 1.18, suggesting a slight nitrogen deficiency but still relatively close to the composition of the as deposited NbN film. As the annealing temperature increased to 950°C, a more pronounced increase in the Nb/N ratio was observed, reaching 1.21 for shorter annealing times (15 seconds), and 1.83 after 30 minutes of annealing. This significant rise in Nb/N ratio at 950°C indicates a notable loss of nitrogen from the film, likely due to nitrogen out-diffusion at elevated temperatures, particularly with prolonged annealing, leading to non-stoichiometric films.

The electrical properties of the NbN films were systematically evaluated using temperature-dependent four-point probe resistivity measurements, which are crucial for assessing the performance of these films in metallic applications, particularly where low resistivity is essential for efficient electrical performance [39]. Figure 5(a) presents the temperature dependence of film resistance, normalized to the resistance at 300 K ($\rho/\rho_{300}$), for the as-deposited and annealed NbN samples. A distinct trend emerges: the as-deposited and 500°C-annealed films show very similar resistivity ratios, indicating that moderate annealing does not significantly alter the resistivity, while the sample annealed at 950°C exhibits a much higher resistivity ratio, especially at lower temperature regions. The similarity in the normalized resistivity between the as-deposited and 500°C-annealed films suggests that the microstructural improvements observed at 500°C, such as strain relaxation and dislocation healing, do not substantially affect the carrier scattering mechanisms and helps maintain a relatively low resistivity across the temperature range, similar to the as-deposited state. In contrast, the 950°C-annealed sample shows a higher resistivity ratio. This behavior can be attributed to the introduction of non-stoichiometry, specifically nitrogen out-diffusion, as confirmed by XPS measurements, which show an increase in the Nb/N ratio after annealing at 950°C. The increased resistivity is likely due to the formation of nitrogen vacancies, which enhance carrier scattering and degrade the electrical performance.

The residual resistivity ratio (RRR), defined as the ratio of resistance at 300 K to resistance at 20 K, provides further insight into the temperature-dependent resistivity behavior. Figure 5(b) shows that the RRR values are relatively stable for both the as-deposited and 500°C-annealed films, remaining around 0.85. This stability suggests that the moderate annealing conditions at 500°C do not introduce additional scattering centers, and thus preserve the film's structural and electrical integrity. However, the 950°C-annealed films exhibit a sharp reduction in RRR, with values ranging from 0.4 to as low as 0.29. This reduction in RRR also reflects film higher order of non-stoichiometry at higher annealing temperatures, contributing to the enhanced scattering of charge carriers. The superconducting critical temperature ($T_c$) of the NbN films, shown in Figure 5(c), mirrors the trends observed in RRR. Both the as-deposited and 500°C-annealed films maintain relatively high $T_c$ values between 12-13 K, indicating that the 500°C annealing temperature preserves the superconducting properties without introducing significant nitrogen loss or structural degradation for all three durations investigated. In contrast, the 950°C-annealed films exhibit a reduction in $T_c$, with values dropping to 8-10 K. This reduction is likely caused by the increased nitrogen vacancies and residual strain within the films, both of which may disrupt the long-range coherence necessary for maintaining the superconducting state by introducing scattering centers which can interfere with electron pairing, thereby degrading the superconducting characteristics of the films. Prior studies on NbTiN films deposited by plasma-enhanced atomic layer deposition (PEALD) on AlN-buffered silicon substrates [40] demonstrated improved superconducting performance using slow thermal annealing (STA, 3.33 °C/min) at 1000 °C compared to rapid thermal annealing (RTA, 60 °C/min), with both conducted in a nitrogen atmosphere, largely due to improved crystallinity and reduced disorder. However, the response to thermal processing can differ between NbTiN and NbN systems, particularly due to differences in material composition, deposition technique, annealing atmosphere, ramp rates and durations, and substrate selection, highlighting the importance of process-specific optimization when aiming to preserve or enhance superconducting performance.

In Figure 6(a), the relationship between lattice constants and annealing conditions is presented, derived from the (111) δ-NbN XRD peak positions. Table 1 also provides a comprehensive summary of the properties of as-deposited and annealed NbN films under different annealing conditions. Across the as-deposited, 500°C, and 950°C annealed samples, the lattice constant shows distinct variations that correlate with changes in the microstructure and electrical properties. For the as-deposited sample, the lattice constant remains relatively large as shown in Figure 6(a), which can be attributed to the presence of a high density of dislocations and internal strain, as confirmed by the STEM analysis. The as-deposited films exhibit higher order of structural disorder, likely due to strain accumulated during the growth. Annealing at 500°C results in a reduction of the lattice constant, with an improvement in crystallographic coherence with improved surface morphologies as compared to the as-deposited sample. This annealing condition also preserves similar superconducting properties, as shown in Figure 6(b). The reduction in lattice constant suggests that the annealing process at this temperature facilitates atomic rearrangement and strain relaxation without significant nitrogen loss, maintaining the film's structural and electrical integrity. After 950°C annealing, the lattice constant decreases slightly; along with the nitrogen out-diffusion, evidenced by a significant increase in the Nb/N ratio to 1.83 after 30 minutes of annealing, leads to a significant reduction in $T_c$ (down to 8.69 K) and higher electrical resistivity, as reflected in the lower RRR values (as low as 0.29). Such reduction in $T_c$ with decreasing lattice constants has also been observed in sputtered NbN films grown on AlN substrates at various growth temperatures [26].

Scanning transmission electron microscopy was employed to elucidate the microstructural characteristics of $\delta$-NbN thin films. Figure 7 presents the cross-sectional STEM-HAADF micrographs of the as-deposited and annealed $\delta$-NbN films. Both crystals were projected along $[\bar{1}10]_{NbN}//[100]_{GaN}$ direction with the nominal $[111]_{NbN}//[001]_{GaN}$ orientation upwards. In the as-deposited film, Figure 7(a) reveals a high density of threading dislocations suggesting a strain-relaxation mechanism inherent to the epitaxial growth process. These dislocations are a direct consequence of the lattice mismatch between the NbN film and the GaN substrate, which induces localized strain that is subsequently relieved through defect formation. The observation of fewer dislocations near the NbN/GaN interface could be justified by the Matthews-Blakeslee model [41, 42], where strain energy first builds up within the film and only leads to dislocation formation at greater thicknesses. As the film thickness increases beyond a critical threshold, the strain becomes energetically unfavorable, resulting in the formation of misfit dislocations to relieve this excess strain. Alternatively, the reduced prominence of these dislocations near the NbN/GaN interface may be attributed to the employed imaging orientation, where certain components of dislocation loops are not visible due to geometric factors. Figure 7(b) provides a higher magnification view, highlighting the coexistence of disordered and relatively ordered regions near the NbN/GaN interface, indicative of complex interplay between the need to accommodate internal stresses and the simultaneous effort to maintain crystallographic integrity throughout the thin film. The NbN/GaN interface appears relatively indistinct, likely due to a significant lattice misfit strain at the heteroepitaxial junction, which contributes to the observed microstructural irregularities.

Upon thermal annealing, substantial changes in the microstructure are observed. The atomic structure of 950 °C-annealed (for 1 mins) $\delta$-NbN was characterized in Figure 7(c)-(f). As shown in Figure 7(c), the annealed $\delta$-NbN film exhibits a noticeable change in defect appearance, compared to its as-deposited counterpart, highlighting the critical role of annealing process in facilitating dislocation mobility and atomic rearrangement. The reduced defects observed at atomic scale suggest that the thermal treatment helps recrystallization, resulting in better uniformity in the crystalline structure. This refinement is further evident in Figure 7(d), where the annealed film exhibits a higher degree of atomic order, and the NbN/GaN interface appears sharper and better defined than in the as-deposited film, which can be justified by the more compact atomic arrangement in the annealed sample, as corroborated quantitatively by XRD results in Figure 6. Additionally, the reduced RMS roughness of the sample observed post-annealing (Figure 3(f)) further supports the positive impact of thermal treatment on the quality of the NbN film. These observations imply that thermal annealing serves as a relaxation mechanism, partially relieving the lattice misfit strain between the δ-NbN film and GaN substrate. This strain relaxation promotes the formation of a more thermodynamically stable crystal structure. The observed contrast between the enhanced cross-sectional STEM images, particularly at the NbN/GaN interface regions, and the broader XRD rocking curve FWHM values, as shown in Figure 1(b) for the 950 °C-annealed films, can be attributed to dislocation dynamics at the interface. During annealing, a buildup of misfit dislocations, including sessile 90° (Lomer-type) and glissile 60° dislocations, likely occurs to relieve lattice mismatch strain [43, 44]. These misfit dislocations, along with their associated threading dislocations, may contribute to strain relaxation, thereby promoting a more stable crystalline structure. However, the rearrangement and interaction of dislocations during this process may introduce localized strain fields, which could explain the observed broadening of XRD rocking curves. At the atomic scale, two distinct crystal arrangements are identified in bulk regions of the annealed δ-NbN film, as illustrated in Figures 7(e) and (f). Careful crystal structure fitting reveals that these regions correspond to growth along [111] [Figure 7($e_1$)] and [11$\bar{1}$] [Figure 7($e_2$)] orientation, respectively. These orientations are further confirmed by the FFT analysis in Figure 7($f_1$) and Figure 7($f_2$). The coexistence of these degenerated orientations is not unexpected, given the intrinsic film growth orientation degeneracy, which points to a complex growth mechanism, potentially governed by local variations in substrate-film interactions and thermal gradients during the annealing process.

Given the resemblance between the atomic structure of tetragonal $\gamma$-$Nb_3N_4$ and cubic $\delta$-NbN when viewed along the same [$\bar{1}10$] direction, further clarification is necessary to resolve the ambiguity in phase identification. Indeed, as shown in Figure 8(a), the atomic arrangements of both tetragonal $\gamma$-$Nb_3N_4$ and cubic $\delta$-NbN exhibit a strong visual similarity, but the tetragonal $\gamma$-$Nb_3N_4$ phase can only exhibit the arrangement akin to our observations when the crystal is oriented such that its upward direction aligns with [112]. However, the simulated electron diffraction pattern [Figure 8(b)] of this if-and-only-if scenario differs from the experimentally observed FFT results. This discrepancy, coupled with our XPS results indicating a Nb:N ratio within the cubic range (albeit off-stoichiometric), allows us to rule out the possibility of the tetragonal $\gamma$-$Nb_3N_4$ phase [45, 46]. We thus confirm that the annealed $\delta$-NbN crystal retains its cubic structure. This detailed microstructural analysis highlights the significant changes in the film's morphology and crystallography as a result of thermal annealing, offering valuable insights into the growth dynamics and thermal stability of $\delta$-NbN thin films on GaN substrates. The evolution of the $\delta$-NbN microstructure from the as-deposited to the annealed state demonstrates the critical role of post-deposition thermal treatment in improving the crystalline quality of the $\delta$-NbN thin films and enhancing the overall structural coherence.

Finally, to further assess the applicability of NbN for space and extreme environment applications, the effects of gamma ($\gamma$) radiation on its crystallinity, lattice structure, and superconducting properties were investigated. The NbN films were exposed to $^{60}$Co gamma-ray irradiation with energies of 1.173 MeV and 1.332 MeV at a rate of 1Mrad (Si)/hr for 5 hours at room temperature, corresponding to a cumulative absorbed dose of 5 Mrad (Si). The results, as summarized in Table 2, provide valuable insight into the radiation tolerance of these films, with several key findings highlighting its resilience in extreme radiation environment. The XRD rocking curve FWHM of the (111) NbN showed a slight decrease from 0.25° in the as-deposited sample to 0.24° after γ-irradiation, suggesting that the crystallinity of the NbN films remained largely unaffected by such high radiation dose. In addition to crystallinity, the lattice constant of the NbN films experienced a minor reduction from 4.43 Å in the as-deposited sample to 4.41 Å post-irradiation. This slight contraction in the lattice parameter could be attributed to the introduction of point defects or the reconfiguration of atomic positions due to radiation-induced damage. Nevertheless, the small magnitude of this change indicates that the overall crystal structure of NbN remains stable under γ-irradiation. Furthermore, the RRR decreased slightly from 0.85 to 0.80 after irradiation, suggesting that radiation-induced defects may have introduced additional scattering centers, slightly increasing the resistivity. The $T_c$ showed a slight decrease from 12.82 K in the as-deposited sample to 12.65 K after irradiation. This negligible drop in $T_c$ highlights the robustness of the superconducting phase in NbN under γ-irradiation, reinforcing the material's potential for use in superconducting applications in space, where maintaining superconductivity in the presence of radiation is critical. Table 2 also compares the properties of NbN films pre- and post-radiation. Overall, the NbN films demonstrate high resilience to gamma radiation, with only minimal changes observed in crystallinity, lattice constant, RRR, and superconducting temperature, making it a promising material for space applications and other radiation-intensive environments.

## IV. CONCLUSION

In summary, this study provides a comprehensive analysis of how high temperature annealing and high dose gamma radiation affects the structural, morphological, and superconducting properties of NbN films deposited on GaN. The annealing process was found to significantly impact film quality and superconducting characteristics. XRD confirmed that the cubic δ-NbN films grow epitaxially on GaN, aligning their lattice orientation with the c-axis of the substrate. Annealing at moderate temperatures improved crystallinity, reduced lattice strain, and enhanced surface smoothness, as evidenced by narrower rocking curve FWHMs and reduced surface RMS roughness. Higher annealing temperatures for extended duration, however, induced nitrogen out-diffusion and led to increased mosaicity and roughness, correlating with degradation in superconducting and electrical properties. XPS revealed nitrogen deficiency in films annealed at elevated temperatures, contributing to stoichiometric imbalances and enhanced carrier scattering, which increased resistivity and lowered the residual resistivity ratio. Despite these changes, STEM analysis confirmed that the films retained their cubic δ-NbN structure, with no evidence of phase transformation. The NbN films exhibited a complex microstructural evolution, characterized by the presence of dislocations and rotational domains, which were influenced by the substrate-film interactions and thermal annealing conditions. Additionally, the NbN films demonstrated exceptional radiation resilience, maintaining their crystallinity, lattice structure, and superconducting properties even after exposure to high-dose gamma radiation. These findings offer valuable insights into the annealing-modulated characteristics of NbN superconductors and the influence of radiation on their properties, fostering the development of robust superconducting devices integrated with III-nitride semiconductor technologies for multifunctional applications in quantum computing, space, defense and other extreme environments.

## ACKNOWLEDGMENT

B.L, G.C.O., and J.H acknowledge funding support from National Science Foundation under DMR-1847964. The authors also acknowledge the start-up funding support provided by University of Massachusetts Lowell, including support from the Provost's Office and the Vice Chancellor of Research. S. M. recognizes additional support from Department of the Air Force under Air Force Contract No. FA8702-15-D-0001. Any opinions, findings, conclusions or recommendations expressed in this material are those of the author(s) and do not necessarily reflect the views of the Department of the Air Force. Delivered to the U.S. Government with Unlimited Rights, as defined in DFARS Part 252.227-7013 or 7014 (Feb 2014). Notwithstanding any copyright notice, U.S. Government rights in this work are defined by DFARS 252.227-7013 or DFARS 252.227-7014 as detailed above. Use of this work other than as specifically authorized by the U.S. Government may violate any copyrights that exist in this work.

## CONFLICT OF INTEREST

The authors have no conflicts to disclose.

## DATA AVAILABILITY

The data that support the findings of this study are available from the corresponding author upon reasonable request.

**Table 1.**

Summary of superconducting transition temperature, compositional stoichiometry, XRD rocking curve FWHMs, lattice constants, surface RMS roughness, and residual resistance ratio for the as deposited and high temperature annealed samples.

| Annealing Temperature (°C) | Annealing Duration (min) | Critical Temperature (K) | Nb/N Ratio | Residual Resistance Ratio ($\rho_{300K}/\rho_{20K}$) | Lattice Constant (Å) | Rocking Curve FWHM (°) | Surface RMS Roughness (nm) |
|---|---|---|---|---|---|---|---|
| As-deposited | - | 12.82 | 1.13 | 0.846 | 4.43 | 0.248 | 1.29 |
| 500 | 0.25 | 12.80 | 1.16 | 0.861 | 4.41 | 0.255 | 1.00 |
| 500 | 1 | 12.85 | 1.18 | 0.907 | 4.41 | 0.188 | 0.63 |
| 500 | 30 | 12.35 | 1.18 | 0.785 | 4.40 | 0.210 | 0.60 |
| 950 | 0.25 | 9.31 | 1.21 | 0.403 | 4.37 | 0.800 | 0.85 |
| 950 | 1 | 9.60 | 1.22 | 0.387 | 4.37 | 0.792 | 0.65 |
| 950 | 30 | 8.69 | 1.83 | 0.291 | 4.38 | 1.040 | 2.01 |

**Table 2.**

Comparison of superconducting transition temperatures, lattice constants, residual resistance ratio, XRD rocking curve FWHMs of as deposited and γ-irradiated samples after 5 Mrad (Si) of cumulative dose.

| Sample | (111) Cubic NbN FWHM (°) | Lattice Constant (Å) | Residual Resistance Ratio | Critical Temperature (K) |
|---|---|---|---|---|
| As-deposited | 0.25 | 4.43 | 0.85 | 12.82 |
| γ-irradiated | 0.24 | 4.41 | 0.80 | 12.65 |

**Figure Captions**

**Figure 1.** (a) XRD ω-2θ scan profiles for the (111) reflection of δ-NbN films deposited on GaN. (b) XRD rocking curve FWHM values of (111) δ-NbN based on annealing conditioning. (c) XRD Φ-scan profiles of the asymmetric peak of (105) GaN and (311) NbN for the annealed (950˚C, 1 minute) sample.

**Figure 2.** X-ray Asymmetrical reciprocal space mapping of (240) δ-NbN and (114) GaN for (a) as-deposited, (b) 500°C (30 minute), and (c) 950°C (30 minute) annealed samples.

**Figure 3.** AFM images (5μm x 5μm scan area) of δ-NbN films grown on GaN: (a) as-deposited and annealed at (b-d) 500°C and (e-g) 900°C for annealing duration of (b, e) 15 sec, (c, f) 1 min, and (d, g) 30 mins.

**Figure 4.** Nb/N ratio determined from Nb 3s and N 1s core level peak areas using XPS for as-deposited and annealed samples.

**Figure 5.** (a) Resistance vs Temperature profile for as-deposited and annealed δ-NbN films. (b) Residual resistance ratio and (c) Superconducting transition temperature ($T_c$) for NbN films under different annealing conditions.

**Figure 6.** (a) Lattice constants (Å) calculated from (111) δ-NbN XRD peak positions for NbN films under different annealing conditions. (b) Superconducting transition temperature ($T_c$) vs Lattice constant (Å) of NbN films.

**Figure 7.** Cross-sectional HAADF micrographs of δ-NbN thin films: (a-b) as-deposited and (c-f) annealed (950 °C for 1 min). Threading dislocations and the nominal crystallographic orientations of δ-NbN film and GaN substrate are indicated in (a) and (c). (b) and (d) Higher magnification

view of the as-deposited and annealed films at the interface. (e₁) and (f₁) Atomic resolution micrographs of two degenerate growth orientations superimposed with fitted crystal structures. (e₂) and (f₂) Indexed FFT patterns of (e₁) and (f₁), respectively.

**Figure 8.** Crystal structure model (a) and simulated diffraction pattern (b) of the $\gamma$-Nb$_4$N$_3$ crystal oriented to [$\bar{1}$10]-zone axis.

**Figure 1.**

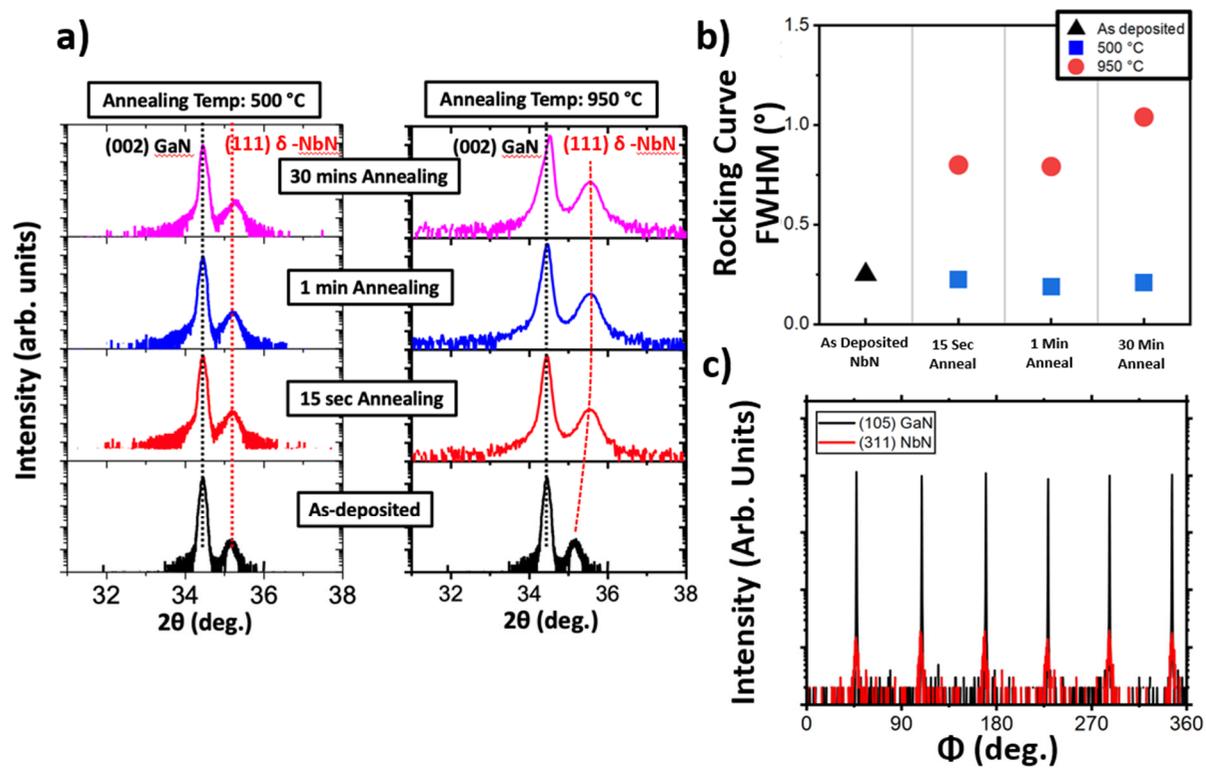

**Figure 2.**

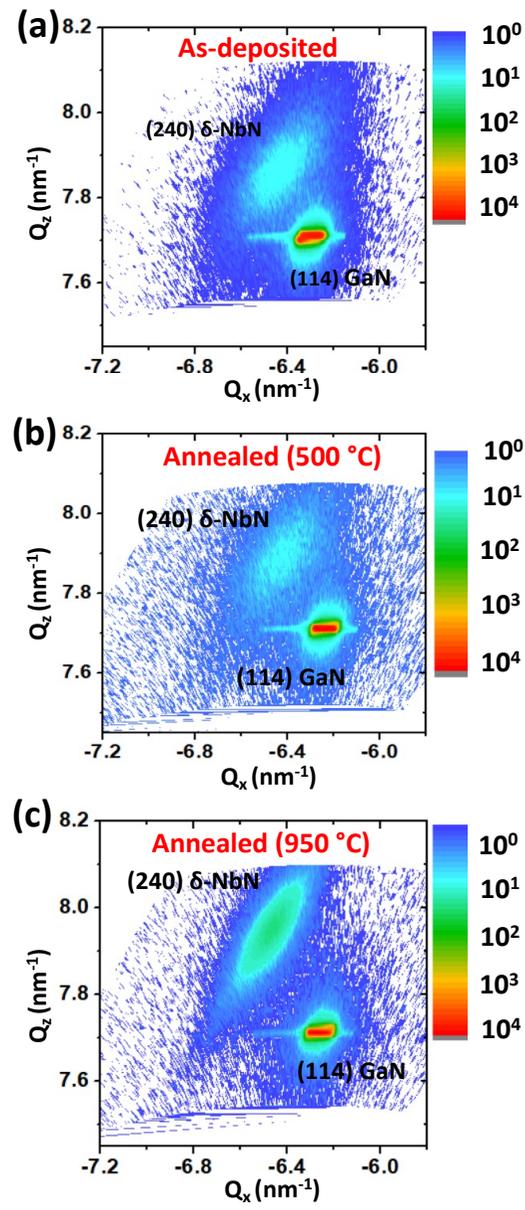

**Figure 3.**

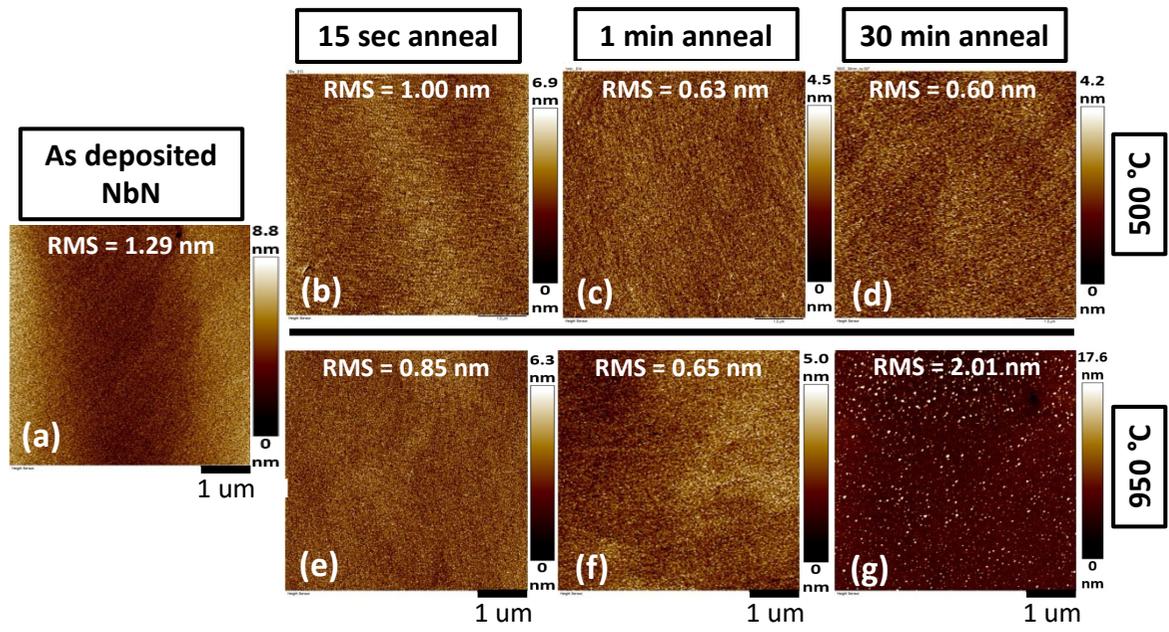

**Figure 4.**

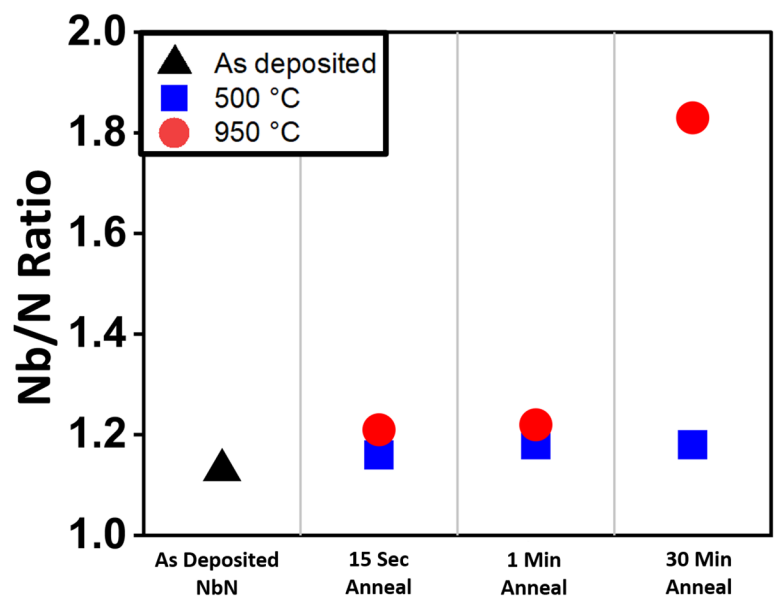

**Figure 5.**

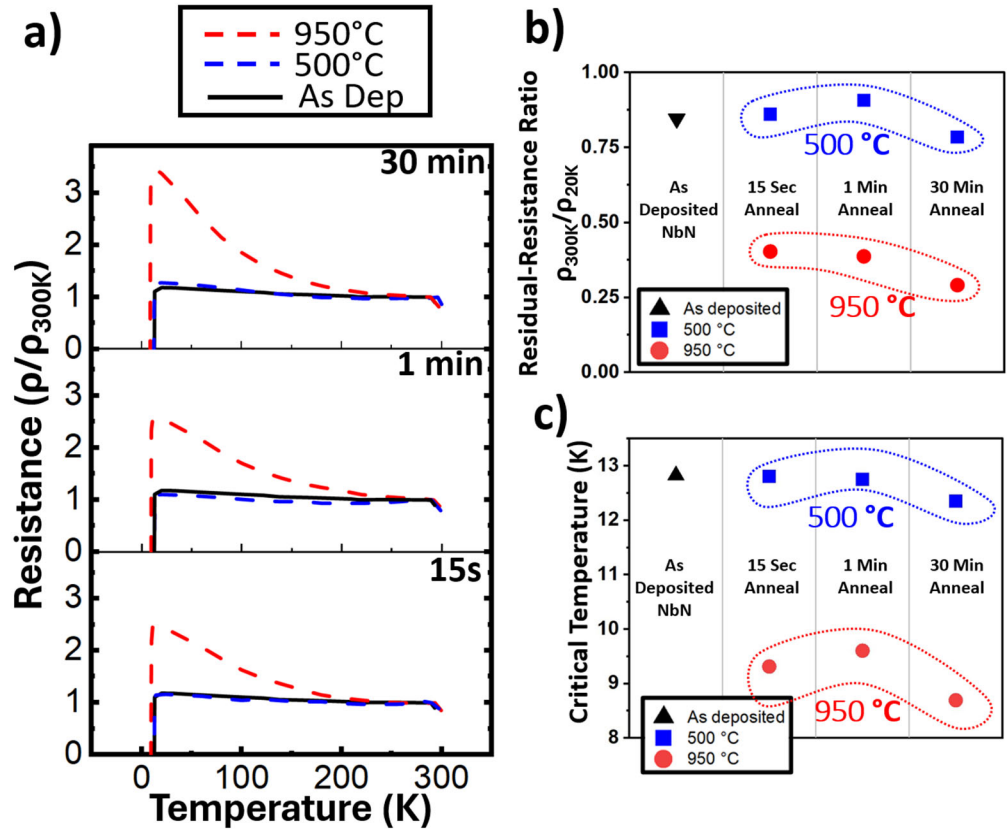

**Figure 6.**

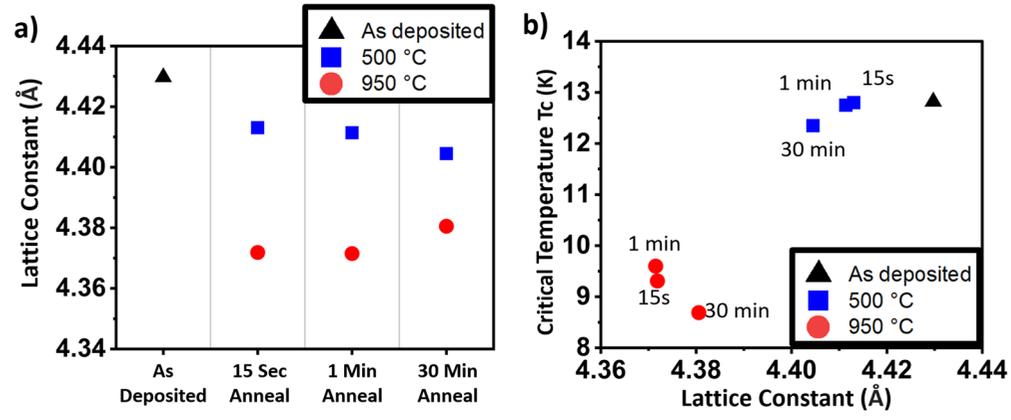

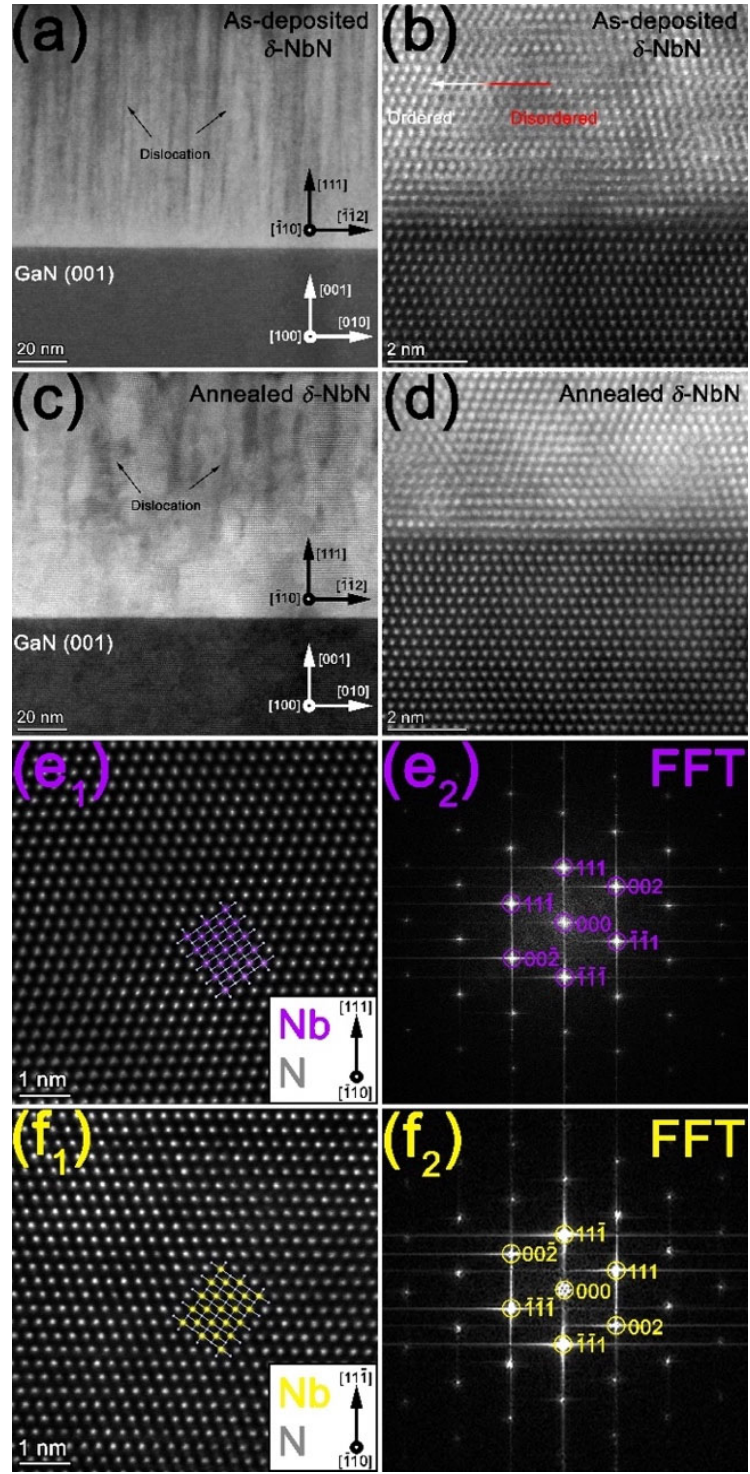

**Figure 8.**

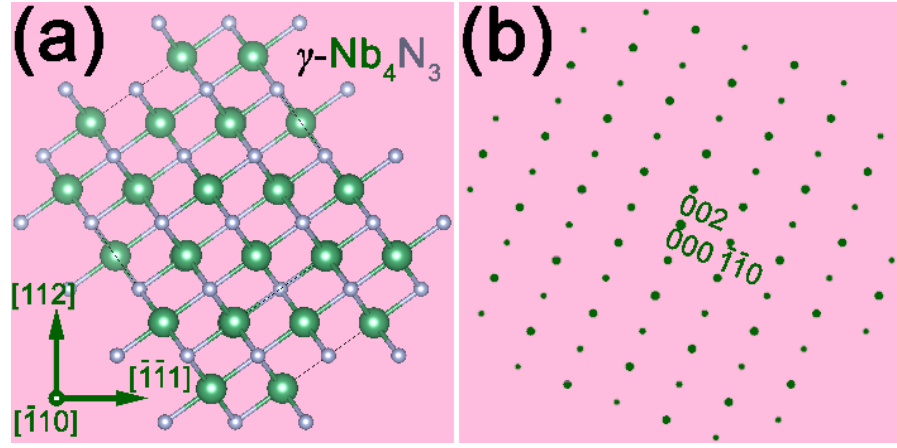